\documentclass[aps,twocolumn,superscriptaddress,amsmath,amssymb]{revtex4}

\usepackage{graphicx}
\usepackage{dcolumn}
\usepackage{bm}

\begin{document}

\title{Decomposition mechanism and the effects of metal additives on the kinetics of lithium alanate}

\author{Khang Hoang}
\affiliation{Materials Department, University of California, Santa Barbara, California 93106, USA}%
\affiliation{Computational Materials Science Center, George Mason University, Fairfax, Virginia 22030, USA}%
\author{Anderson Janotti}
\affiliation{Materials Department, University of California, Santa Barbara, California 93106, USA}%
\author{Chris G. Van de Walle}
\affiliation{Materials Department, University of California, Santa Barbara, California 93106, USA}%


\begin{abstract}

First-principles density functional theory studies have been carried out for native defects and transition-metal (Ti and Ni) impurities in lithium alanate (LiAlH$_{4}$), a potential material for hydrogen storage. On the basis of our detailed analysis of the structure, energetics, and migration of lithium-, aluminum-, and hydrogen-related defects, we propose a specific atomistic mechanism for the decomposition and dehydrogenation of LiAlH$_{4}$ that involves mass transport mediated by native point defects. We also discuss how Ti and Ni impurities alter the Fermi-level position with respect to that in the undoped material, thus changing the concentration of charged defects that are responsible for mass transport.  This mechanism provides an explanation for the experimentally observed lowering of the temperature for the onset of decomposition and of the activation energy for hydrogen desorption from LiAlH$_{4}$.

\end{abstract}

\maketitle

\section{\label{sec:intro}Introduction}

Lithium alanate (LiAlH$_{4}$) has been considered as a potential material for hydrogen storage due to its high hydrogen density and relatively low decomposition temperature.~\cite{orimo_chem_rev_2007} Yet the atomistic mechanisms behind the decomposition and dehydrogenation processes in this complex hydride are far from understood. For the purpose of optimizing its hydrogen storage and release capacity, it is desirable to understand the rate-limiting processes involved in the hydrogen desorption. First-principles calculations based on density functional theory have been demonstrated to be powerful for addressing defect-related processes in solids, and have provided valuable insights in the atomistic mechanisms involved in mass transport and hydrogen release.~\cite{peles_prb_2007,hoang_prb_2009,wilson-short_prb_2009,hoang_angew} Here we apply this approach to explore possible mechanisms for the decomposition and dehydrogenation of LiAlH$_{4}$.

It has been observed that LiAlH$_{4}$ desorbs hydrogen through a two-step process, similar to that reported in the widely studied NaAlH$_{4}$, i.e.,~\cite{dilts_1972,dymova_1994,blanchard_2005,andreasen_2005}
\begin{equation}
\label{eq:step1}
\rm{LiAlH_{4} \rightarrow \frac{1}{3} Li_{3}AlH_{6} + \frac{2}{3}Al + H_{2}},
\end{equation}
\begin{equation}
\label{eq:step2}
\rm{\frac{1}{3}Li_{3}AlH_{6} \rightarrow LiH + \frac{1}{3}Al + \frac{1}{2}H_{2}}.
\end{equation}
The first reaction occurs around 112$-$220$^\circ$C with a theoretical hydrogen release of 5.3 wt\%. It was observed to initiate by the melting of LiAlH$_{4}$ in the temperature range 150$-$170$^\circ$C, although isothermal decomposition without melting has also been reported.~\cite{andreasen_2005,wiench_2004,varin_2010} The second reaction takes place around 127$-$260$^\circ$C and releases 2.6 wt\% hydrogen. On the other hand, using nuclear magnetic resonance (NMR) studies of the decomposed sample during isothermal heating of LiAlH$_{4}$ at 150$^\circ$C over 2h, Wiench {\it et al}.~\cite{wiench_2004} observed an apparent deviation of the sample composition from that predicted by the above two-step mechanism. They suggested that thermal decomposition of LiAlH$_{4}$ may follow several different reaction paths, including
\begin{equation}
\label{eq:alternative}
\rm{LiAlH_{4} \rightarrow LiH + Al + \frac{3}{2}H_{2}}.
\end{equation}
Later studies carried out by Varin {\it et al}.,~\cite{varin_2010} however, did not support such direct decomposition of LiAlH$_{4}$ in the solid state into LiH and Al.

Significant efforts have been devoted to studying the decomposition kinetics of LiAlH$_{4}$ and determining the activation energy for hydrogen desorption.~\cite{andreasen_2005,andreasen_2006,blanchard_2005,varin_2010} From kinetic measurements carried out under isothermal conditions, Andreasen {\it et al}.~\cite{andreasen_2005} obtained an apparent activation energy of 0.85 eV for eqn (\ref{eq:step1}) in the solid state. For the same reaction but in the liquid state, Andreasen obtained an activation energy of 0.84 eV.~\cite{andreasen_2006} Blanchard {\it et al}.~\cite{blanchard_2005} reported an activation energy of 1.06 eV for the main desorption stage of LiAlD$_{4}$. More recently, Varin and Zbroniec estimated the activation energy for eqn (\ref{eq:step1}) to be 1.15 eV and 0.96 eV for as-received and ball-milled LiAlH$_{4}$, respectively.~\cite{varin_2010} Ball milling thus results in a slightly lower activation energy.

It has also been reported that metal additives such as Ti and Ni improve the dehydrogenation properties of LiAlH$_{4}$.~\cite{blanchard_2004,blanchard_2005,wang_adsorption_2005,chen_jpc_2001,kojima_2008,liu_jacs_2009,sun_ijhe_2008,andreasen_2006,naik_ijhe_2009,langmi_jpc_2010,rafiuddin_2010,varin2010928,varin20111167} For example, ball-milling LiAlH$_{4}$ with NiCl$_{2}$ was found to reduce the onset decomposition temperature by about 50$^\circ$C.~\cite{sun_ijhe_2008} Doping with TiCl$_{3}$ was also found to lower the onset temperature of eqn (\ref{eq:step1}) by 60$-$75$^\circ$C, bringing it well below the melting point of LiAlH$_{4}$.~\cite{langmi_jpc_2010} Recently, Liu {\it et al}.~\cite{liu_jacs_2009} demonstrated that Ti-doped LiAlH$_{4}$ can operate as a reversible hydrogen storage material that can release up to 7 wt\% hydrogen commencing at temperatures as low as 80$^\circ$C, and recharge can be achieved by employing liquid dimethyl ether as a solvent. In another study, nanometric TiC and Ni additives have been found to reduce the effective decomposition temperature and enhance the hydrogen desorption of LiAlH$_{4}$.\cite{rafiuddin_2010,varin2010928,varin20111167}

In spite of the consensus among different experimental reports that Ti and Ni lower the onset decomposition temperature, the conclusions regarding the effects of metal additives on the kinetics of isothermal decomposition of LiAlH$_{4}$ are not without conflict. On the one hand, Andreasen~\cite{andreasen_2006} reported apparent activation energies of 0.84 and 0.92 eV for undoped and Ti-doped LiAlH$_{4}$, respectively. Since the differences in the activation energies between undoped and Ti-doped samples were within the experimental uncertainty, the author suggested that the effect of Ti doping on the dehydrogenation kinetics of LiAlH$_{4}$ was mainly on the prefactor.~\cite{andreasen_2006} Blanchard {\it et al}.\cite{blanchard_2005} also reported small differences between the activation energies of doped and undoped samples; e.g., 1.06 and 0.99 eV for undoped and TiCl$_{3}$$\cdot$1/3AlCl$_{3}$-doped LiAlD$_{4}$, respectively. On the other hand, Chen {\it et al}.~\cite{chen_jpc_2001} found that the activation energy for hydrogen desorption of LiAlH$_{4}$ doped with 2 mol\% TiCl$_{3}$$\cdot$1/3AlCl$_{3}$ is 0.44 eV for eqn (\ref{eq:step1}) in the solid state, which is much smaller than the reported values (0.84$-$1.15 eV) for undoped LiAlH$_{4}$. Other research groups reported smaller, but still significant, reductions in the activation energy for decomposition when the compound was ball-milled with metal additives.~\cite{rafiuddin_2010,varin2010928} For example, Varin {\it et al}.\cite{varin2010928} reported an activation energy of 0.73 eV for LiAlH$_{4}$ ball-milled with nanometric Ni, which is 0.23 eV lower than that for undoped LiAlH$_{4}$.\cite{varin_2010}

To resolve this situation, clearly one needs to understand the fundamental mechanisms behind the decomposition and dehydrogenation processes and the interaction between the metal additives and the host material. Theoretical studies have so far focused mainly on bulk LiAlH$_{4}$ and its thermodynamic properties.~\cite{lovvik_prb_2004,kang_jpc_2004,vajeeston_2005,yoshino_2005,vansetten_prb_2007,jang_2006} Experimental data, on the other hand, suggested that the decomposition process involves mass transport by native point defects.~\cite{andreasen_2005,Ares20081263} This motivates us to perform first-principles calculations of native defects and transition-metal (Ti and Ni) impurities in LiAlH$_{4}$. As we will discuss in this paper, these calculations enable us to explore possible mechanisms for the decomposition and dehydrogenation involving mass transport mediated by native point defects and investigate the effects of Ti and Ni impurities in the material.

\section{\label{sec;metho}Methodology}

{\bf Computational details.} Our calculations were based on density functional theory within the generalized-gradient approximation (GGA)~\cite{GGA} and the projector augmented wave method,~\cite{PAW1,PAW2} as implemented in the VASP code.~\cite{VASP1,VASP2,VASP3} Calculations for bulk LiAlH$_{4}$ (24 atoms/unit cell) were performed using a 10$\times$6$\times$6 Monkhorst-Pack $\mathbf{k}$-point mesh.~\cite{monkhorst-pack} For calculations of native defects and transition-metal impurities, we used a (2$\times$2$\times$2) supercell containing 192 atoms/cell, and a 2$\times$2$\times$2 $\mathbf{k}$-point mesh. The plane-wave basis-set cutoff was set to 400 eV and convergence with respect to self-consistent iterations was assumed when the total energy difference between cycles was less than 10$^{-4}$ eV and the residual forces were less than 0.01 eV/{\AA}. The migration of selected native defects in LiAlH$_{4}$ was studied using the climbing image nudged elastic band method (NEB).~\cite{ci-neb}

{\bf Defect formation energies.} The likelihood of forming a defect is given by its formation energy ($E^{f}$). In thermal equilibrium, the concentration of defect X at temperature $T$ can be obtained via the relation~\cite{walle:3851,janotti2009}
\begin{equation}\label{eq:concen}
c(\mathrm{X})=N_{\mathrm{sites}}N_{\mathrm{config}}\mathrm{exp}[-E^{f}(\mathrm{X})/k_BT],
\end{equation}
where $N_{\mathrm{sites}}$ is the number of high-symmetry sites in the lattice per unit volume on which the defect can be incorporated, and $N_{\mathrm{config}}$ is the number of equivalent configurations per site. Obviously, defects with lower formation energies are more likely to form and occur in higher concentrations. Note that the energy in eqn (\ref{eq:concen}) is, in principle, a free energy; however, the entropy and volume terms are often neglected because they are negligible at relevant experimental conditions.\cite{janotti2009}

The formation energy of a defect X in charge state $q$ is defined as~\cite{walle:3851}
\begin{equation}\label{eq:eform}
E^f({\mathrm{X}}^q)=E_{\mathrm{tot}}({\mathrm{X}}^q)-E_{\mathrm{tot}}({\mathrm{bulk}})-\sum_{i}{n_i\mu_i} +q(E_{\mathrm{v}}+\Delta V+\mu_{e}),
\end{equation}
where $E_{\mathrm{tot}}(\mathrm{X}^{q})$ and $E_{\mathrm{tot}}(\mathrm{bulk})$ are, respectively, the total energies of a supercell containing the defect X and of a supercell of the perfect bulk material. $\mu_{i}$ is the chemical potential of species $i$; $\mu_{i}$=$\mu_{i}^{0}$+$\tilde{\mu}_{i}$, where $\mu_{i}^{0}$ equals the chemical potential of element $i$ in its standard state. $n_{i}$ denotes the number of atoms of species $i$ that have been added ($n_{i}$$>$0) or removed ($n_{i}$$<$0) to form the defect. $\mu_{e}$ is the electron chemical potential, i.e., the Fermi level, referenced to the valence-band maximum in the bulk ($E_{\mathrm{v}}$). $\Delta V$ is the ``potential alignment'' term, i.e., the shift in the band positions due to the presence of the charged defect and the neutralizing background, obtained by aligning the average electrostatic potential in regions far away from the defect to the bulk value.~\cite{walle:3851}

{\bf Chemical potentials.} The atomic chemical potentials $\mu_{i}$ are variables and can be chosen to represent experimental conditions. In the following discussions, we assume that LiAlH$_{4}$, Al, and Li$_{3}$AlH$_{6}$ are stable and in equilibrium. The chemical potentials of Li, Al, and H can then be obtained from the equations that express the stability of LiAlH$_{4}$, Al, and Li$_{3}$AlH$_{6}$.~\cite{walle:3851} This gives, approximately, $\tilde{\mu}_{\rm Li}$=$-$0.862 eV, $\tilde{\mu}_{\rm Al}$=0 eV, and $\tilde{\mu}_{\rm H}$=0 eV. Note that this set of chemical potentials also approximately corresponds to assuming that LiAlH$_{4}$, H$_{2}$, and Al (or Li$_{3}$AlH$_{6}$) are in equilibrium. For the impurities (Ti and Ni), the chemical potentials are fixed to the energy of the bulk metals, $\tilde{\mu}_{i}$=0 eV, which is the upper bound.

{\bf Transition levels.} We will refer to $\epsilon(q_{1}/q_{2})$ as the thermodynamic transition level which is defined as the Fermi-level position where the charge states $q_{1}$ and $q_{2}$ have equal formation energies.~\cite{walle:3851} It can be shown from eqn (\ref{eq:eform}) that the Fermi level at which the transition takes place is
\begin{equation}\label{eq:translevel}
\epsilon(q_{1}/q_{2})=\frac{E_{\mathrm{tot}}({\mathrm{X}}^{q_{1}})-E_{\mathrm{tot}}({\mathrm{X}}^{q_{2}})+(q_{1}-q_{2})E_{\mathrm{v}}}{q_{2}-q_{1}},
\end{equation}
where $q_{1}$ and $q_{2}$ are the initial and final charge states, respectively. Clearly, $\epsilon(q_{1}/q_{2})$ is independent of the choice of atomic chemical potentials.

\section{\label{sec;bulk}Bulk properties}

\begin{figure}
\begin{center}
\includegraphics[width=2.2in]{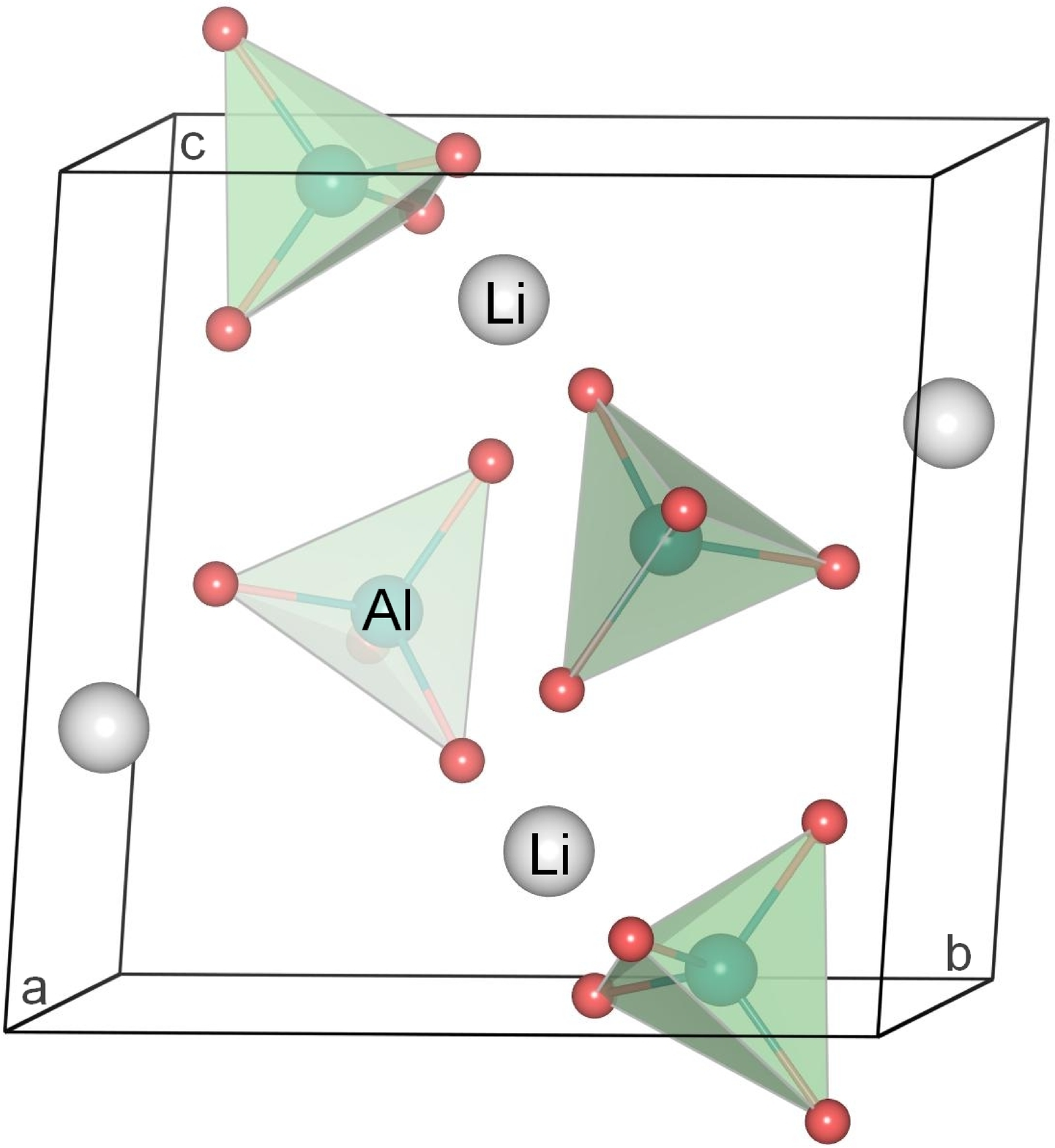}
\end{center}
\vspace{-0.15in}
\caption{Relaxed structure of monoclinic LiAlH$_{4}$. Large (gray) spheres are Li, medium (blue) spheres Al, and small (red) spheres H.}\label{fig;struct}
\end{figure}

We start by presenting the results for the basic structural and electronic properties of bulk LiAlH$_{4}$. The compound was reported to crystallize in the monoclinic structure, space group $P\mathrm{2}_{\mathrm{1}}/c$, with lattice parameters $a$=4.817 {\AA}, $b$=7.802 {\AA}, $c$=7.821 {\AA}, and $\beta$=112.228$^{\circ}$ at 8 K.~\cite{hauback_jac_2002} It can be regarded as an ordered arrangement of Li$^{+}$ and (AlH$_{4}$)$^{-}$ units. Figure \ref{fig;struct} shows the optimized structure of LiAlH$_{4}$. The calculated lattice parameters $a$=4.860 {\AA}, $b$=7.817 {\AA}, $c$=7.832 {\AA}, and $\beta$=111.808$^{\circ}$ are in agreement with the experimental values.

\begin{figure}
\begin{center}
\includegraphics[width=3.4in]{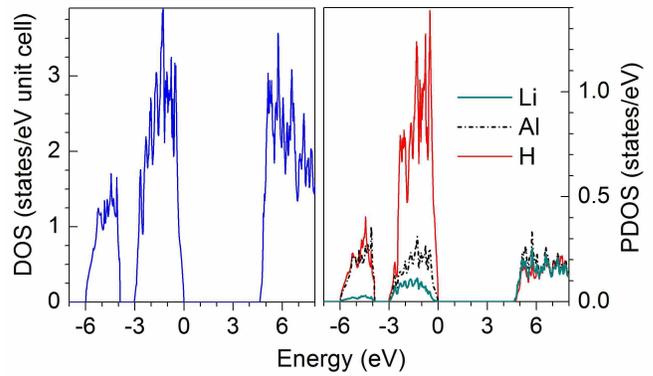}
\end{center}
\vspace{-0.15in}
\caption{Total density of states (DOS) and projected density of states (PDOS) of LiAlH$_{4}$. The zero of energy is set to the highest occupied state.}\label{fig;DOS}
\end{figure}

Figure \ref{fig;DOS} shows the total density of states and projected density of states of LiAlH$_{4}$. The valence-band maximum (VBM) consists of the bonding state of Al $p$ and H $s$, whereas the conduction-band minimum (CBM) consists of the antibonding state of Al $p$ and H $s$ and contribution from Li $s$. The calculated band gap is 4.64 eV, very close to that reported previously (4.67 eV).~\cite{vansetten_prb_2007} As we will illustrate in the next sections, knowing the structural and electronic properties of LiAlH$_{4}$ is essential to understand the properties of native defects and the interaction between impurities and the host compound.

\section{\label{sec:defects}Formation of native defects}

In insulating, wide band-gap materials such as LiAlH$_{4}$, native point defects are expected to exist in charged states other than neutral, and charge neutrality requires that defects with opposite charge states coexist in equal concentrations.~\cite{peles_prb_2007,hoang_prb_2009,wilson-short_prb_2009,hoang_angew} We therefore investigated hydrogen-, lithium-, and aluminum-related point defects in all possible charge states. Defect complexes are also considered, with special attention to Frenkel pairs, i.e., interstitial-vacancy pairs of the same species. In the following, we present the results for the defects in each category. The role of these defects in ionic and mass transport in LiAlH$_{4}$ will be discussed in the next section.

\begin{figure}
\begin{center}
\includegraphics[width=3.0in]{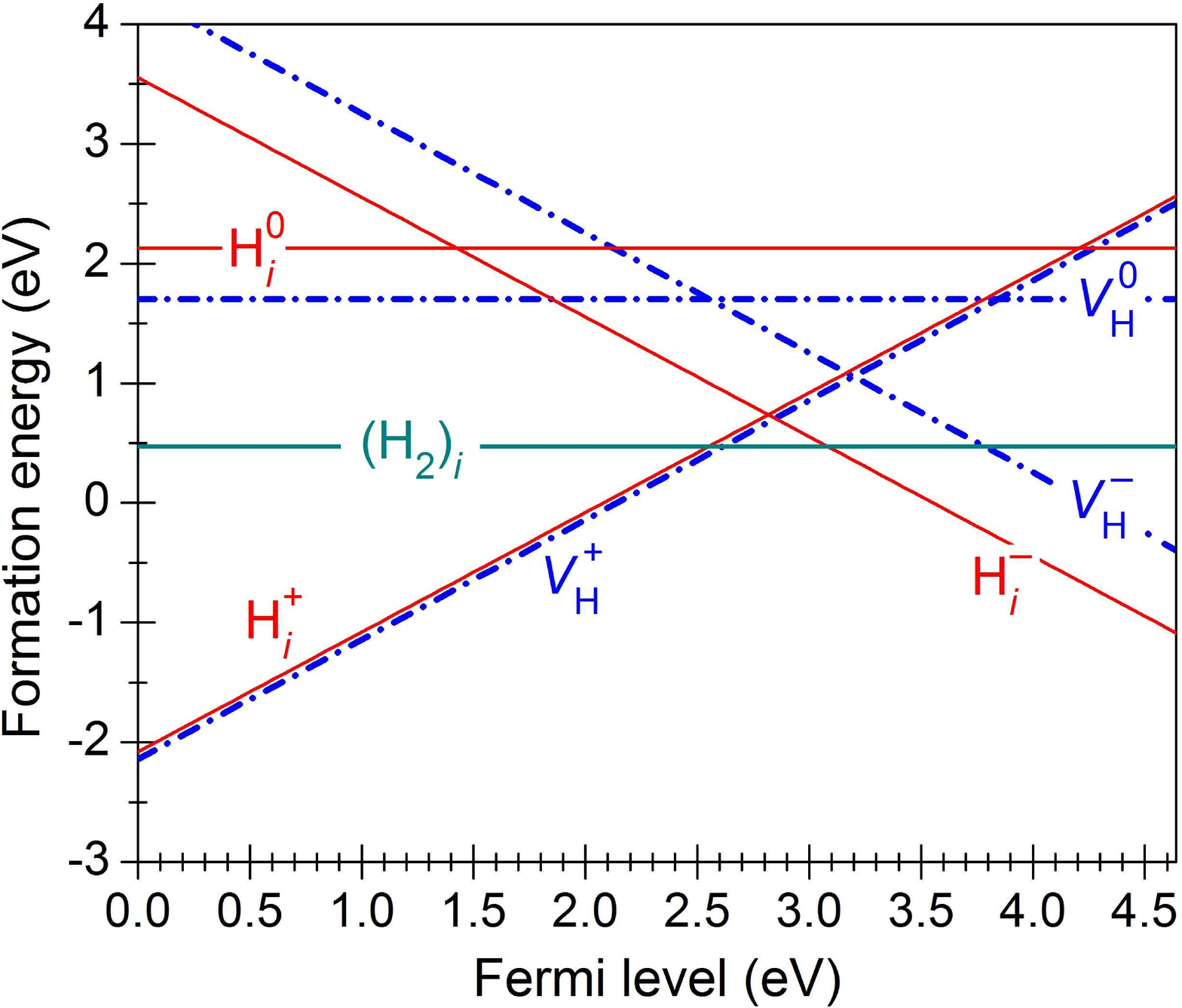}
\end{center}
\vspace{-0.15in}
\caption{Calculated formation energies of hydrogen-related defects in LiAlH$_{4}$, plotted as a function of Fermi level with respect to the valence-band maximum.}\label{fig;FE;H}
\end{figure}

{\bf Hydrogen-related defects.} Figure \ref{fig;FE;H} shows the calculated formation energies of hydrogen vacancies ($V_{\mathrm{H}}$), interstitials (H$_{i}$), and interstitial molecules (H$_{2}$)$_{i}$ in LiAlH$_{4}$. Among these defects, the positively charged hydrogen vacancy ($V_{\mathrm{H}}^{+}$), positively charged hydrogen interstitial (H$_{i}^{+}$), and negatively charged hydrogen interstitial (H$_{i}^{-}$) have the lowest formation energies over a wide range of Fermi-level values. (H$_{2}$)$_{i}$ has the lowest formation energy in a relatively small range near $\mu_{e}$=2.84 eV where the formation energies of $V_{\mathrm{H}}^{+}$ and H$_{i}^{-}$ are equal. The neutral hydrogen vacancy ($V_{\mathrm{H}}^{0}$) and interstitial (H$_{i}^{0}$) are energetically less favorable than their respective charged defects over the entire range of the Fermi-level values, which is a characteristic of negative-$U$ centers.~\cite{negativeU}

\begin{figure}
\begin{center}
\includegraphics[width=3.2in]{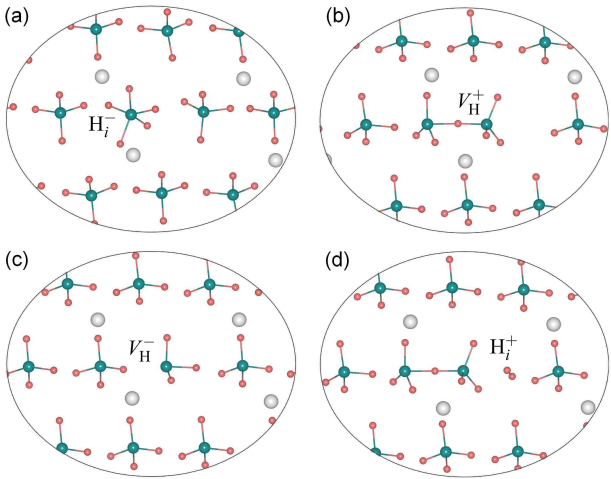}
\end{center}
\vspace{-0.15in}
\caption{Structures of (a) H$_{i}^{-}$, (b) $V_{\mathrm{H}}^{+}$, (c) $V_{\mathrm{H}}^{-}$, and (d) H$_{i}^{+}$. Only Li and Al atoms in [010] planes (and their coordinated H atoms) are shown.}\label{fig;H}
\end{figure}

The creation of H$_{i}^{-}$ involves adding one H atom and an electron (i.e., H$^{-}$) to the LiAlH$_{4}$ supercell (hereafter referred to as ``the system''). This H$^{-}$ combines with an (AlH$_{4}$)$^{-}$ unit to form (AlH$_{5}$)$^{2-}$, see Fig.~\ref{fig;H}(a). In the (AlH$_{5}$)$^{2-}$, the average Al$-$H distance is 1.71 {\AA}, compared to 1.63 {\AA} in the perfect bulk crystal. One may expect that due to Coulomb interaction, H$_{i}^{-}$ would prefer to stay near a Li$^{+}$ unit. In fact, we find that this configuration is not stable, except in some cases where H$_{i}^{-}$ is created simultaneously with other defect(s) (which will be illustrated later when we present our results for aluminum-related defects).

$V_{\mathrm{H}}^{+}$ is created by removing H$^{-}$ from the system. This leads to formation of an AlH$_{3}$-H-AlH$_{3}$ complex, or (Al$_{2}$H$_{7}$)$^{-}$, with the Al$-$Al distance being of 3.26 {\AA} (compared to 3.99 {\AA} in the bulk), see Fig.~\ref{fig;H}(b). $V_{\mathrm{H}}^{-}$, on the other hand, can be thought as the extraction of an H$^{+}$ ion from the system. This results in an AlH$_{3}$ unit, see Fig.~\ref{fig;H}(c). (H$_{2}$)$_{i}$ involves adding an H$_{2}$ molecule to the supercell. This interstitial molecule prefers to stay in an interstitial void, with the calculated bond length of 0.75 {\AA} being equal to that calculated for an isolated H$_{2}$ molecule. Finally, H$_{i}^{+}$ is created by adding an H$^{+}$ ion into the system. This results in an AlH$_{3}$-H-AlH$_{3}$ complex plus a H$_{2}$ interstitial molecule, see Fig.~\ref{fig;H}(d). H$_{i}^{+}$ can therefore be regarded as a complex of $V_{\mathrm{H}}^{+}$ and (H$_{2}$)$_{i}$. The formation energy of H$_{i}^{+}$ is, however, lower than the sum of the formation energies of $V_{\mathrm{H}}^{+}$ and (H$_{2}$)$_{i}$, giving H$_{i}^{+}$ a binding energy of 0.44 eV with respect to its constituents. From these analyses, it is evident that H$_{i}^{-}$, $V_{\rm{H}}^{+}$, $V_{\rm{H}}^{-}$, and (H$_{2}$)$_{i}$ are elementary native point defects, meaning the structure and energetics of the other defects can be interpreted in terms of these basic building blocks.

For the migration of H$_{i}^{-}$, $V_{\rm{H}}^{+}$, $V_{\rm{H}}^{-}$, and (H$_{2}$)$_{i}$, we find energy barriers of 0.15, 0.63, 0.90, and 0.23 eV, respectively. The energy barriers for $V_{\rm{H}}^{+}$ and $V_{\rm{H}}^{-}$ are relatively high because the diffusion of these defects involves breaking Al$-$H bonds from AlH$_{4}$ units. The diffusion of $V_{\rm{H}}^{-}$, for example, involves moving an H atom from an AlH$_{4}$ unit to the vacancy. The saddle-point configuration in this case consists of a H atom located midway between two AlH$_{3}$ units (i.e., AlH$_{3}$-H-AlH$_{3}$). Such a configuration is favorable in the case of $V_{\mathrm{H}}^{+}$, but high in energy for $V_{\rm{H}}^{-}$. For H$_{i}^{+}$, which is a complex of $V_{\mathrm{H}}^{+}$ and (H$_{2}$)$_{i}$, the migration barrier is larger than or equal to that of the least mobile constituent,\cite{wilson-short_prb_2009} i.e., 0.63 eV, the value for $V_{\mathrm{H}}^{+}$.

\begin{figure}
\begin{center}
\includegraphics[width=2.4in]{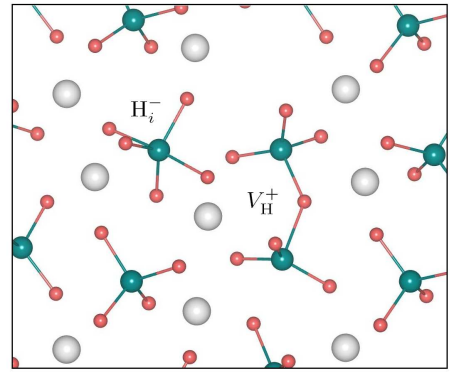}
\end{center}
\vspace{-0.15in}
\caption{Structure of (H$_{i}^{-}$,$V_{\mathrm{H}}^{+}$) Frenkel pair in LiAlH$_{4}$. Only Li and Al atoms in [100] planes (and their coordinated H atoms) are shown. }\label{fig;Frenkel;H}
\end{figure}

Considering that hydrogen vacancies and interstitials can be stable as oppositely charged defects, charge and mass conservation conditions suggest that these native defects may form in the interior of the material in the form of Frenkel pairs. Since in LiAlH$_{4}$, H$_{i}^{+}$ is a complex defect, the only possible hydrogen Frenkel pair is (H$_{i}^{-}$,$V_{\mathrm{H}}^{+}$), whose structure is shown in Fig.~\ref{fig;Frenkel;H}. The configurations of the individual defects are preserved in this complex. (H$_{i}^{-}$,$V_{\mathrm{H}}^{+}$) has a formation energy of 1.23 eV, and a binding energy of 0.20 eV with respect to isolated H$_{i}^{-}$ and $V_{\mathrm{H}}^{+}$.

\begin{figure}
\begin{center}
\includegraphics[width=3.0in]{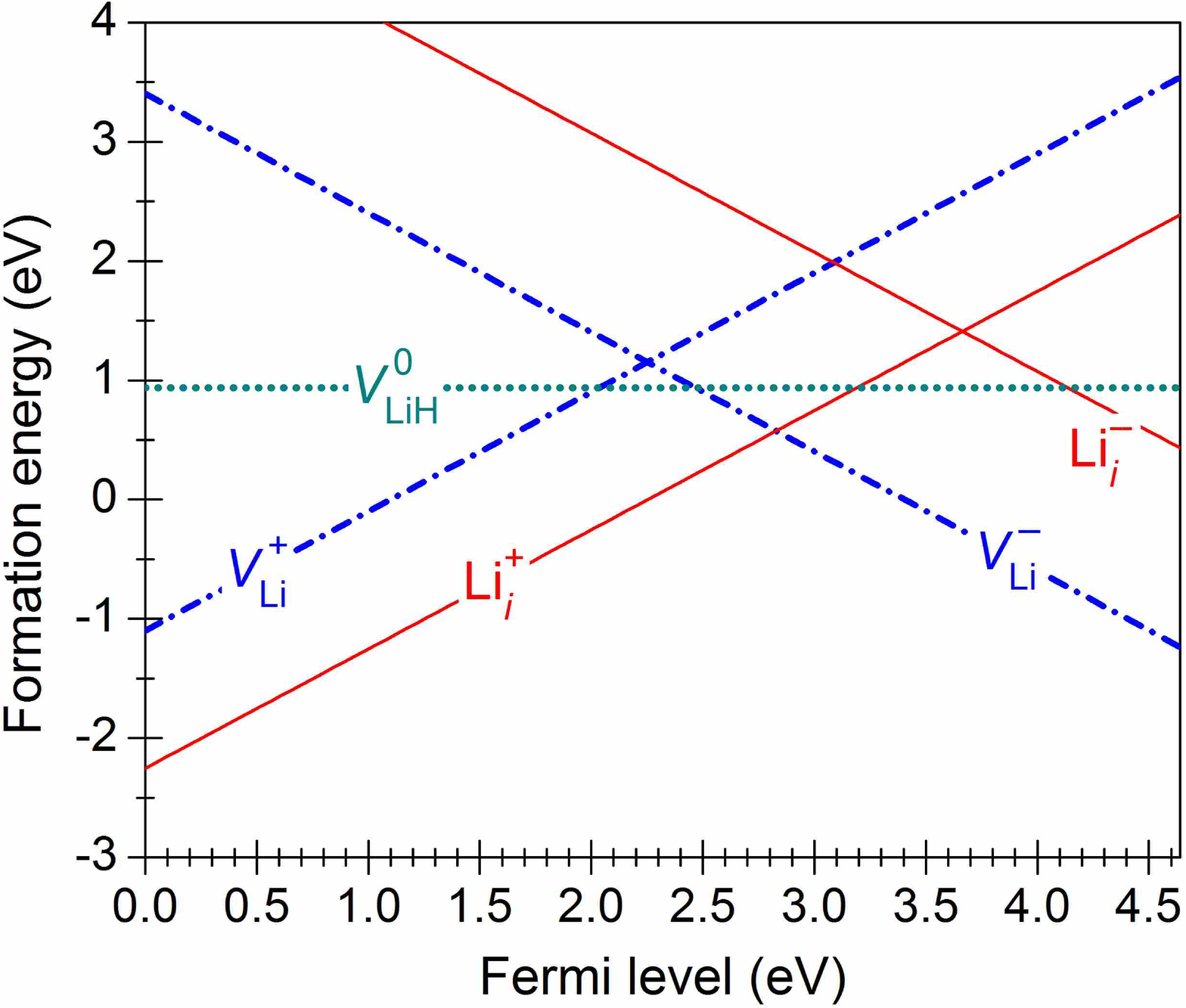}
\end{center}
\vspace{-0.15in}
\caption{Calculated formation energies of lithium-related defects in LiAlH$_{4}$, plotted as a function of Fermi level with respect to the valence-band maximum.}\label{fig;FE;Li} 
\end{figure}

{\bf Lithium-related defects.} Figure \ref{fig;FE;Li} shows the calculated formation energies of lithium vacancies ($V_{\mathrm{Li}}$), interstitials (Li$_{i}$), and $V_{\mathrm{LiH}}^{0}$ (removing one Li and one H atom) in LiAlH$_{4}$. Among these lithium-related defects, Li$_{i}^{+}$ and $V_{\mathrm{Li}}^{-}$ have the lowest formation energies for the entire range of Fermi-level values. These two defects have equal formation energies at $\mu_{e}$=2.83 eV.

The creation of $V_{\mathrm{Li}}^{-}$ involves removing a Li$^{+}$ ion from the system. This causes very small changes to the lattice geometry. On contrary, $V_{\mathrm{Li}}^{+}$, created by removing a Li atom and an extra electron, strongly disturbs the system. Besides the void formed by the removed Li, there are two AlH$_{3}$-H-AlH$_{3}$ complexes and a H$_{2}$ interstitial molecule that all can be identified as 2$V_{\mathrm{H}}^{+}$ and (H$_{2}$)$_{i}$. Therefore, $V_{\mathrm{Li}}^{+}$ can be regarded as a complex of $V_{\mathrm{Li}}^{-}$, 2$V_{\mathrm{H}}^{+}$, and (H$_{2}$)$_{i}$.

Li$_{i}^{+}$ is created by adding a Li$^{+}$ ion to the system. Like $V_{\mathrm{Li}}^{-}$, Li$_{i}^{+}$ does not cause much disturbance to the lattice geometry. On the other hand, Li$_{i}^{-}$, which is created by adding a Li atom and an extra electron to the system, strongly disturbs the system by breaking Al$-$H bonds and forming AlH$_{3}$ and AlH$_{5}$ units which can be identified as $V_{\mathrm{H}}^{-}$ and H$_{i}^{-}$, respectively. This defect, therefore, is considered as a complex of Li$_{i}^{+}$, $V_{\mathrm{H}}^{-}$, and H$_{i}^{-}$. Similarly, $V_{\mathrm{LiH}}^{0}$ can be regarded as a complex of $V_{\mathrm{Li}}^{-}$ and $V_{\mathrm{H}}^{+}$. Thus, Li$_{i}^{+}$ and $V_{\mathrm{Li}}^{-}$ can be considered as the elementary defects in the Li sublattice.

The migration of Li$_{i}^{+}$ involves an energy barrier of 0.28 eV. For $V_{\mathrm{Li}}^{-}$, the migration involves moving Li$^{+}$ from a nearby lattice site to the vacancy, and this gives an energy barrier as low as 0.14 eV. The migration barriers for Li$_{i}^{+}$ and $V_{\mathrm{Li}}^{-}$ are relatively small, suggesting that they are highly mobile. For Li$_{i}^{-}$, which can be considered as a complex of Li$_{i}^{+}$, $V_{\mathrm{H}}^{-}$, and H$_{i}^{-}$, the migration barrier is estimated to be at least 0.90 eV, the value for $V_{\mathrm{H}}^{-}$. Similarly, the estimated migration barrier of $V_{\mathrm{Li}}^{+}$ and $V_{\mathrm{LiH}}^{0}$ is at least 0.63 eV, the value for $V_{\mathrm{H}}^{+}$.

We also investigated possible formation of lithium Frenkel pairs. Since Li$_{i}^{-}$ and $V_{\mathrm{Li}}^{+}$ are not elementary defects, the only possibility is (Li$_{i}^{+}$,$V_{\mathrm{Li}}^{-}$). The distance between Li$_{i}^{+}$ and $V_{\mathrm{Li}}^{-}$ is 3.64 {\AA}. This pair has a formation energy of 0.75 eV and a binding energy of 0.40 eV. The formation energy is, therefore, much lower than that of the hydrogen Frenkel pair (H$_{i}^{-}$,$V_{\mathrm{H}}^{+}$). This result indicates that LiAlH$_{4}$ may be prone to Frenkel disorder on the Li sublattice.

\begin{figure}
\begin{center}
\includegraphics[width=3.0in]{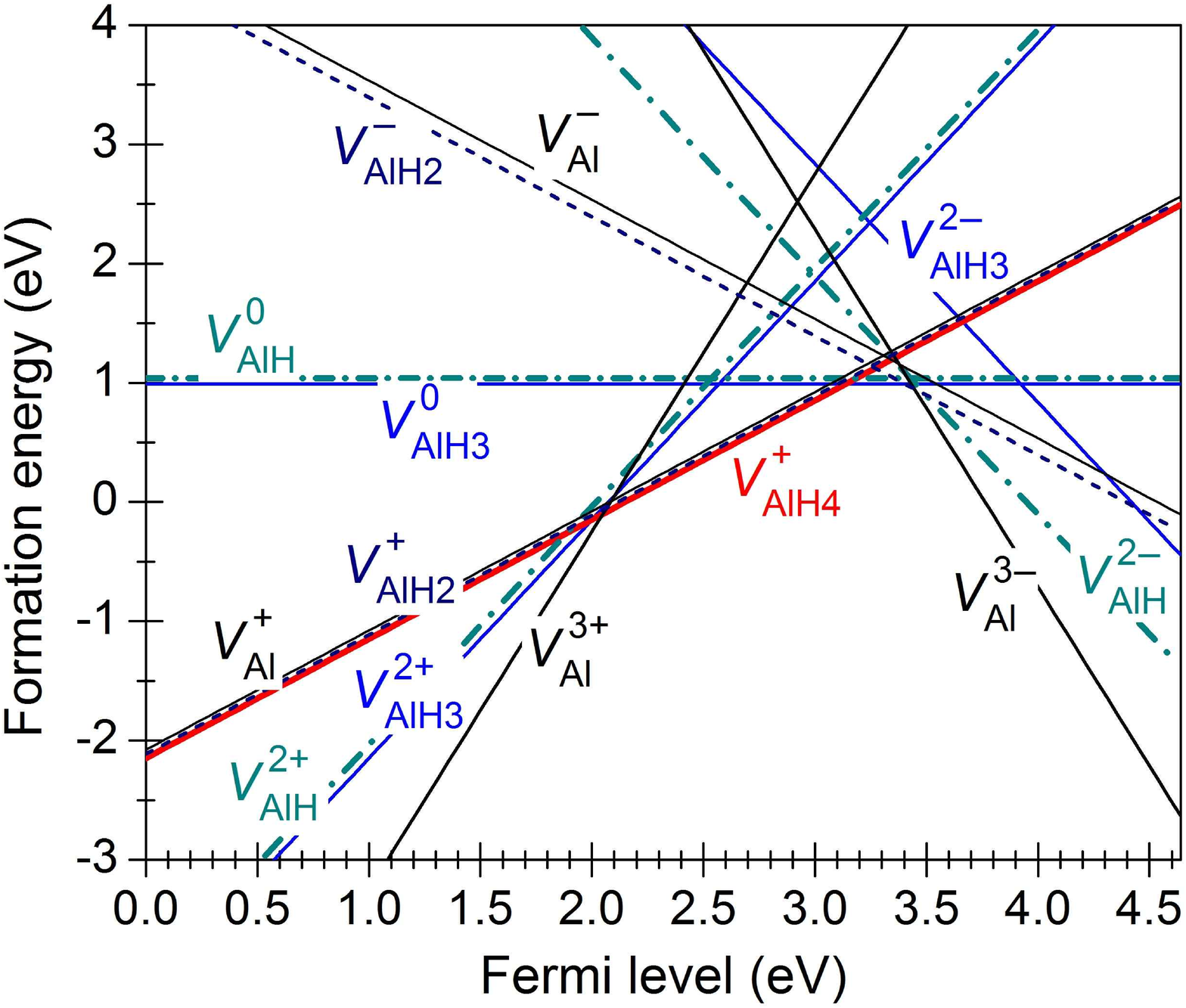}
\end{center}
\vspace{-0.15in}
\caption{Calculated formation energies of aluminum-related defects in LiAlH$_{4}$, plotted as a function of Fermi level with respect to the valence-band maximum.}\label{fig;FE;Al}
\end{figure}

{\bf Aluminum-related defects.} Figure \ref{fig;FE;Al} shows the calculated formation energies of Al vacancies ($V_{\mathrm{Al}}$), AlH vacancies ($V_{\mathrm{AlH}}$), AlH$_{2}$ vacancies ($V_{\mathrm{AlH_{2}}}$), AlH$_{3}$ vacancies ($V_{\mathrm{AlH_{3}}}$), and AlH$_{4}$ vacancies ($V_{\mathrm{AlH_{4}}}$). Only the lowest energy charge states for each defect are included. We find that $V_{\mathrm{AlH_{3}}}^{0}$, $V_{\mathrm{AlH_{4}}}^{+}$, $V_{\mathrm{Al}}^{3-}$, and $V_{\mathrm{Al}}^{3+}$ have the lowest formation energies for certain ranges of Fermi-level values. $V_{\mathrm{AlH_{4}}}^{+}$ corresponds to the removal of an entire (AlH$_{4}$)$^{-}$ unit from the system. The migration of $V_{\mathrm{AlH_{4}}}^{+}$ involves moving a nearby (AlH$_{4}$)$^{-}$ unit to the vacancy with an energy barrier of 0.43 eV.

\begin{figure}
\begin{center}
\includegraphics[width=3.0in]{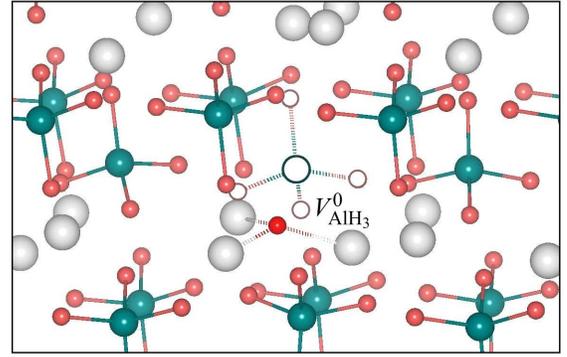}
\end{center}
\vspace{-0.15in}
\caption{Structure of $V_{\mathrm{AlH}_{3}}^{0}$ in LiAlH$_{4}$. The defect can be regarded as a complex of $V_{\mathrm{AlH}_{4}}^{+}$ (presented by empty spheres) and H$_{i}^{-}$. }\label{fig;VAlH3}
\end{figure}

The creation of $V_{\mathrm{AlH_{3}}}^{0}$ involves removing one Al and three H atoms from the system. This is equivalent to removing an (AlH$_{4}$)$^{-}$ unit and adding an H$^{-}$ simultaneously, as shown in Fig.~\ref{fig;VAlH3}. $V_{\mathrm{AlH_{3}}}^{0}$ therefore can be regarded as a complex of $V_{\mathrm{AlH_{4}}}^{+}$ and H$_{i}^{-}$ with a binding energy of 0.42 eV with respect to its constituents. The structure of H$_{i}^{-}$ in this complex is, however, significantly different from that presented earlier, {\it cf.}~Fig.~\ref{fig;H}(a). We find that the H$^{-}$ stays near three Li$^{+}$ ions with the Li$-$H distances being 1.80, 1.84, and 1.85 {\AA}; see Fig.~\ref{fig;VAlH3}. This configuration of the H$_{i}^{-}$ component can play an important role in the formation of LiH. In this case, a certain amount of LiAlH$_{4}$ may decompose directly into LiH, Al, and H$_{2}$ following eqn (\ref{eq:alternative}). The migration barrier of $V_{\mathrm{AlH_{3}}}^{0}$ is estimated to be at least 0.43 eV, given by that of $V_{\mathrm{AlH_{4}}}^{+}$.

We also find an alternative configuration of $V_{\mathrm{AlH_{3}}}^{0}$ where the H$^{-}$ unit combines with an (AlH$_{4}$)$^{-}$ unit to form (AlH$_{5}$)$^{2-}$, which is similar to the local structure of H$_{i}^{-}$ shown in Fig.~\ref{fig;H}(a). This configuration is 0.16 eV higher in energy than the lowest configuration of $V_{\mathrm{AlH_{3}}}^{0}$ mentioned above. Our results thus indicate that H$_{i}^{-}$ prefers staying close to Li$^{+}$ when it is created simultaneously with and in the vicinity of other defects such as $V_{\mathrm{AlH_{4}}}^{+}$.

Similar to $V_{\mathrm{AlH_{3}}}^{0}$, the structure and energetics of other aluminum-related defects can also be interpreted in terms of $V_{\mathrm{AlH_{4}}}^{+}$ and the elementary hydrogen-related defects. $V_{\mathrm{AlH}}^{0}$, for example, can be considered as a complex of $V_{\mathrm{AlH_{4}}}^{+}$, H$_{i}^{-}$, and (H$_{2}$)$_{i}$. The H$_{i}^{-}$ component in this defect is also composed of H$^{-}$ staying close to three Li$^{+}$ ions with the Li$-$H distances being 1.82, 1.85, and 1.85 {\AA}. $V_{\mathrm{AlH_{2}}}^{+}$, on the other hand, can be regarded as a complex of $V_{\mathrm{AlH_{4}}}^{+}$ and (H$_{2}$)$_{i}$. Likewise, $V_{\mathrm{Al}}^{+}$ can be considered as a complex of $V_{\mathrm{AlH_{4}}}^{+}$ and 2(H$_{2}$)$_{i}$; $V_{\mathrm{Al}}^{3-}$ as a complex of $V_{\mathrm{AlH_{4}}}^{+}$, (H$_{2}$)$_{i}$, and 2H$_{i}^{+}$; and $V_{\mathrm{Al}}^{3+}$ as a complex of $V_{\mathrm{AlH_{4}}}^{+}$ and 4H$_{i}^{-}$.

Overall, we find that H$_{i}^-$, $V_{\rm{H}}^-$, $V_{\rm{H}}^+$, (H$_{2})_{i}$, Li$_{i}^+$, $V_{\rm{Li}}^-$, and $V_{\rm{AlH_4}}^+$ are the elementary defects, and that the other defects can be regarded as complexes involving these basic constituents. Understanding the structure and energetics of these native point defects is, therefore, key to describing ionic and mass transport and the decomposition process in LiAlH$_{4}$. Since these point defects, except (H$_{2})_{i}$, are charged, their creation in the interior of the material necessarily requires both mass and charge conservation. As discussed earlier, hydrogen- and lithium-related defects can be formed inside LiAlH$_{4}$ via Frenkel pair mechanisms, i.e., moving H (or Li) from one lattice site to an interstitial site; this leads to formation of interstitial-vacancy complexes such as (H$_{i}^{-}$,$V_{\mathrm{H}}^{+}$) and (Li$_{i}^{+}$,$V_{\mathrm{Li}}^{-}$). Aluminum-related defects such as $V_{\mathrm{AlH_{4}}}^{+}$ and $V_{\mathrm{AlH_{3}}}^{0}$, on the other hand, can only be created at the surface or interface since the creation of such defects inside the material requires creation of the corresponding aluminum-related interstitials which are too high in energy.

\section{\label{sec:transport}Ionic and mass transport}

In the absence of electrically active impurities that can affect the Fermi-level position, or when such impurities occur in much lower concentrations than charged native defects, the Fermi-level position of LiAlH$_{4}$ is determined by oppositely charged defects with the lowest formation energies.~\cite{peles_prb_2007,hoang_prb_2009,wilson-short_prb_2009,hoang_angew} These defects are Li$_{i}^{+}$ and $V_{\rm{Li}}^{-}$ which pin the Fermi level at $\mu_{e}$=2.83 eV (hereafter referred to as $\mu_{e}^{\rm int}$, the Fermi-level position determined by intrinsic/native defects), where the formation energies and hence, approximately, concentrations of Li$_{i}^{+}$ and $V_{\rm{Li}}^{-}$ are equal. We list in Table \ref{tab:lialh} formation energies and migration barriers of the most relevant native point defects and defect complexes in LiAlH$_{4}$. The formation energies for charged defects are taken at $\mu_{e}^{\rm int}$.

\begin{table}[t]
\caption{Formation energies ($E^{f}$) and migration barriers ($E_{m}$) for native defects in LiAlH$_{4}$. Migration barriers denoted by an asterisk ($^{\ast}$) are the lower bounds, estimated by considering the defect as a complex and taking the highest of the migration barriers of the constituents.}\label{tab:lialh}
\begin{center}
\begin{tabular}{cccc}
Defect&$E^f$ (eV)& $E_m$ (eV)&Constituents \\
\hline
H$_{i}^{+}$ &0.73&0.63$^{\ast}$ &$V_{\rm{H}}^{+}$+(H$_{2}$)$_{i}$\\
H$_{i}^{-}$ &0.72&0.15 &	\\
$V_{\rm{H}}^{+}$ &0.70&0.63 & \\
$V_{\rm{H}}^{-}$ &1.42&0.90 \\
(H$_{2}$)$_{i}$&0.47&0.23	\\
Li$_{i}^{+}$	&0.58&0.28	\\
$V_{\rm{Li}}^{-}$&0.58&0.14 \\ 
$V_{\rm{AlH_{4}}}^{+}$&0.68&0.43	\\
$V_{\rm{AlH_{3}}}^{0}$&0.99&0.43$^{\ast}$&$V_{\rm{AlH}_{4}}^{+}$+H$_{i}^{-}$	\\
$V_{\rm{AlH}_{2}}^{+}$&0.69&0.43$^{\ast}$&$V_{\rm{AlH}_{4}}^{+}$+(H$_{2}$)$_{i}$\\
$V_{\rm{AlH}}^{0}$&1.02&0.43$^{\ast}$&$V_{\rm{AlH}_{4}}^{+}$+(H$_{2}$)$_{i}$+H$_{i}^{-}$	\\
$V_{\rm{Al}}^{+}$&0.71&0.43$^{\ast}$&$V_{\rm{AlH}_{4}}^{+}$+2(H$_{2}$)$_{i}$\\
\end{tabular}
\end{center}
\begin{flushleft}
\end{flushleft}
\end{table}

{\bf Lithium-ion conduction.} We note from Table \ref{tab:lialh} that Li$_{i}^{+}$ and $V_{\rm{Li}}^{-}$ have low formation energies and are highly mobile. Both defects can contribute to the ionic conductivity. However, since the calculated migration barrier of $V_{\rm{Li}}^{-}$ is lower than that of Li$_{i}^{+}$, we expect that, in LiAlH$_{4}$, lithium diffusion via vacancy mechanism is dominant. The activation energy for ionic conduction is estimated to be 0.72 eV, the summation of the formation energy and migration barrier of $V_{\rm{Li}}^{-}$ ({\it cf.}~Table \ref{tab:lialh}). This value is in good agreement with the reported experimental value (0.76 eV).\cite{oguchi:096104}

{\bf Decomposition mechanism.} As suggested by experimental data,~\cite{andreasen_2005,Ares20081263} the decomposition of LiAlH$_{4}$ into Li$_{3}$AlH$_{6}$, Al, and H$_{2}$, i.e., eqn (\ref{eq:step1}),  necessarily involves hydrogen and/or aluminum mass transport in the bulk. Besides, local and global charge neutrality must be maintained while charged defects are migrating. Keeping these considerations in mind, we identify the following native defects as essential to the decomposition process:

First, H$_{i}^{-}$, which is expected to help form Li$_{3}$AlH$_{6}$ and/or LiH as discussed in the previous section. The activation energy for self-diffusion of H$_{i}^{-}$ is 0.87 eV, the summation of its formation energy and migration barrier. For other hydrogen-related charged defects such as H$_{i}^{+}$, $V_{\rm{H}}^{+}$, and $V_{\rm{H}}^{-}$, the activation energies are 1.36, 1.33, and 2.32 eV, respectively. These values are much higher than that reported for undoped LiAlH$_{4}$, suggesting that H$_{i}^{+}$, $V_{\rm{H}}^{+}$, and $V_{\rm{H}}^{-}$ are not the native defects that drive the decomposition process.

Second, $V_{\mathrm{AlH_{4}}}^{+}$, which is needed for the diffusion of aluminum-related species and the formation of Al phase. The activation energy for self-diffusion of $V_{\mathrm{AlH_{4}}}^{+}$ is 1.11 eV. Note that some other aluminum-related defects such as $V_{\mathrm{AlH_{2}}}^{+}$ and $V_{\mathrm{Al}}^{+}$ can also play this role. However, $V_{\mathrm{AlH_{4}}}^{+}$ has a lower activation energy and is thus expected to be dominant. Some others such as $V_{\mathrm{AlH_{3}}}^{0}$ and $V_{\mathrm{AlH}}^{0}$ have formation energies that are higher than that of $V_{\mathrm{AlH_{4}}}^{+}$, except for a small range of Fermi-level values above $\mu_{e}$, {\it cf.}~Fig.~\ref{fig;FE;Al}. The activation energies for the diffusion of $V_{\mathrm{AlH_{3}}}^{0}$ and $V_{\mathrm{AlH}}^{0}$ are at least 1.42 and 1.45 eV, respectively, which are also much higher than the experimental values.

Third, Li$_{i}^{+}$ and $V_{\mathrm{Li}}^{-}$, which can be created in the interior of the material in the form of a (Li$_{i}^{+}$,$V_{\mathrm{Li}}^{-}$) Frenkel pair. These low-energy and mobile native point defects can act as accompanying defects in hydrogen/aluminum mass transport, providing local charge neutrality as positively and negatively charged hydrogen- and aluminum-related defects moving in the bulk. They can also participate in mass transport that assists, e.g., the formation of Li$_{3}$AlH$_{6}$ and LiH. The activation energy for the formation and migration of (Li$_{i}^{+}$,$V_{\mathrm{Li}}^{-}$) is estimated to be 0.89 eV, which is the formation energy of the Frenkel pair plus the migration barrier of $V_{\mathrm{Li}}^{-}$.

Given these native defects and their properties, the decomposition of LiAlH$_{4}$ can be described in terms of the following mechanism: $V_{\mathrm{AlH_{4}}}^{+}$ is created at the surface or interface. This is equivalent to removing one (AlH$_{4}$)$^{-}$ unit from the bulk LiAlH$_{4}$. The formation energy of $V_{\mathrm{AlH_{4}}}^{+}$ is relatively low (0.68 eV) in the bulk. This energy is expected to be even lower at the surface, given that the bonding environment at the surface is less constrained than in the bulk. Formation of $V_{\mathrm{AlH_{4}}}^{+}$ is therefore quite likely during decomposition. Since (AlH$_{4}$)$^{-}$ is not stable outside the material, it breaks down into AlH$_{3}$ and H$^{-}$; AlH$_{3}$ subsequently leaves the material and dissociates into Al and H$_{2}$, whereas H$^{-}$ stays at the surface or interface. H$^{-}$ can diffuse into the bulk in form of H$_{i}^{-}$ and combine with an (AlH$_{4}$)$^{-}$ unit to form (AlH$_{5}$)$^{2-}$, an intermediate toward forming (AlH$_{6}$)$^{3-}$, which is essential in forming Li$_{3}$AlH$_{6}$ from LiAlH$_{4}$ according eqn (\ref{eq:step1}). In going from (AlH$_{4}$)$^{-}$ to (AlH$_{6}$)$^{3-}$, the anion unit attracts more and more Li$_{i}^{+}$ due to Coulomb interaction. Here, the highly mobile Li$_{i}^{+}$ will also help maintain local charge neutrality in the region near H$_{i}^{-}$. Note that, instead of forming (AlH$_{5}$)$^{2-}$, some H$_{i}^{-}$ may also be stable in the configuration where the hydrogen interstitial stands near Li$^{+}$ units to form a Li-H complex, {\it cf.}~Fig.~\ref{fig;VAlH3}. This complex may act as a nucleation site for the formation of LiH from LiAlH$_{4}$ according to eqn (\ref{eq:alternative}). In order to maintain the reaction, (AlH$_{4}$)$^{-}$ has to be transported to the surface/interface, which is equivalent to $V_{\mathrm{AlH_{4}}}^{+}$ diffusing into the bulk. As $V_{\mathrm{AlH_{4}}}^{+}$ is migrating, local charge neutrality is maintained by having the highly mobile $V_{\mathrm{Li}}^{-}$ in the vacancy's vicinity.

In the mechanism we just proposed, possible rate-limiting processes are the formation and migration of H$_{i}^{-}$, $V_{\mathrm{AlH_{4}}}^{+}$, and (Li$_{i}^{+}$,$V_{\mathrm{Li}}^{-}$) in the bulk LiAlH$_{4}$. Since $V_{\mathrm{AlH_{4}}}^{+}$ gives the highest activation energy (1.11 eV) among the three defects, with respect to the chosen set of atomic chemical potentials, we believe that the decomposition of LiAlH$_{4}$ is rate-limited by the formation and migration of this defect. The calculated activation energy is in agreement with the reported experimental values (0.84$-$1.15 eV).\cite{andreasen_2005,andreasen_2006,blanchard_2005,varin_2010} Also, because the decomposition and dehydrogenation processes occur at the surface/interface, ball milling that enhances the specific surface area and/or shorten the diffusion paths is expected to slightly enhance the hydrogen desorption kinetics. This is consistent with experimental observations.\cite{andreasen_2005,Ares20081263,varin_2010}

\section{\label{sec:impurities}Transition-metal impurities}

\begin{figure*}
\begin{center}
\includegraphics[width=5.2in]{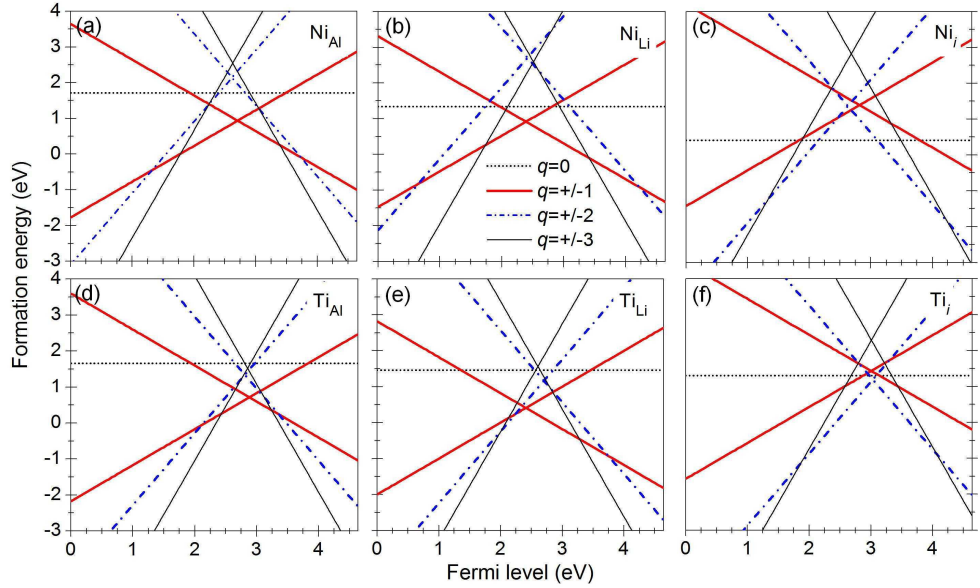}
\end{center}
\vspace{-0.15in}
\caption{Calculated formation energies of transition-metal impurities in different charge states ($q$=0, $\pm$1, $\pm$2, and $\pm$3) in LiAlH$_{4}$, plotted as a function of Fermi level with respect to the valence-band maximum: (a) Ni$_{\mathrm{Al}}$, (b) Ni$_{\mathrm{Li}}$, (c) Ni$_{i}$, (d) Ti$_{\mathrm{Al}}$, (e) Ti$_{\mathrm{Li}}$, and (f) Ti$_{i}$.}\label{fig;FE;TM}
\end{figure*}

We investigate the role played by transition-metal impurities in the hydrogen desorption kinetics of LiAlH$_{4}$ by carrying out calculations for Ti and Ni impurities nominally on the Al and Li sites (i.e., M$_{\mathrm{Al}}$ and M$_{\mathrm{Li}}$, where M= Ti, Ni) and at interstitial sites (M$_{i}$). For each impurity, we performed calculations for several configurations by slightly breaking the symmetry in order to avoid local minima and to obtain the lowest-energy configuration. Figure \ref{fig;FE;TM} shows the calculated formation energies of Ti and Ni impurities in various charge states ($q$=0, $\pm$1, $\pm$2). The results are also summarized in Table \ref{tab:impurities} where we give values for the thermodynamic transition level $\epsilon$(+$q$/$-q$) between charge states $+q$ and $-q$, the associated $U$ value which is defined as $U$=$\epsilon$(0/$-q$)$-$$\epsilon(+q/0)$, the corresponding formation energy $E^{f}$, and the shift in Fermi-level position away from $\mu_{e}^{\rm int}$=2.83 eV (determined by the native defects) which is defined as $\Delta$=$\epsilon$(+$q$/$-$$q$)$-$$\mu_{e}^{\rm int}$. We find that all the substitutional impurities are negative-$U$ centers. Of highest interest are M$_{\mathrm{Al}}^{\pm1}$ and M$_{\mathrm{Li}}^{\pm1}$, which have the lowest formation energies near $\mu_{e}^{\rm int}$=2.83 eV. Regarding the interstitials, Ni$_{i}$ is a positive-$U$ center, whereas Ti$_{i}$ is a negative-$U$ center corresponding to a transition between Ti$_{i}^{2+}$ and Ti$_{i}^{2-}$.

\begin{table}[t]
\caption{Characteristics of Ti and Ni impurities in LiAlH$_{4}$. For Ni$_{i}$ where $U>$0, $E^{f}$ is the formation energy in the neutral charge state (Ni$_{i}^{0}$); this impurity is not effective in shifting the Fermi level and is therefore marked with a $\times$ sign. All the quantities are given in electronvolt (eV).}\label{tab:impurities}
\begin{tabular}{ccrr}
&&Ni&Ti \\
\hline
Al site
&$\epsilon(+/-)$&2.71&2.88 \\
&$U$&$-$1.56&$-$1.90 \\
&$E^{f}$&0.93&0.72 \\
&$\Delta$&$-$0.12&+0.05\\
\hline
Li site
&$\epsilon(+/-)$&2.39&2.41 \\
&$U$&$-$0.85 &$-$2.10 \\
&$E^{f}$&0.92&0.41 \\
&$\Delta$&$-$0.44&$-$0.42 \\
\hline
Interstitial
&$\epsilon$(+2/$-$2)&2.63&3.03 \\
&$U$&+0.95&$-$0.12 \\
&$E^{f}$&0.41&1.19 \\
&$\Delta$&$\times$&+0.20 \\
\end{tabular}

\begin{flushleft}
\end{flushleft}

\end{table}

\begin{figure*}
\begin{center}
\includegraphics[width=5.0in]{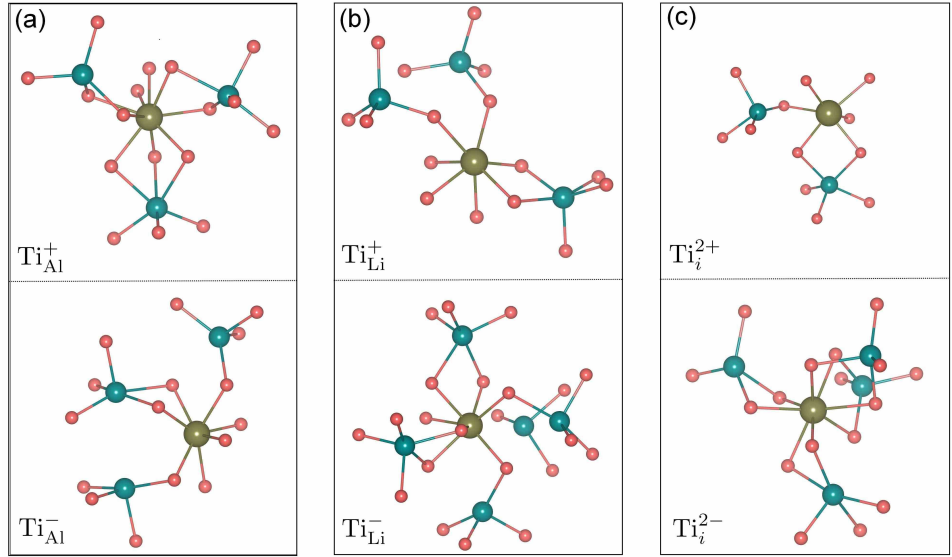}
\end{center}
\vspace{-0.15in}
\caption{Local structures of several substitutional and interstitial impurities in LiAlH$_{4}$: (a) Ti$_{\mathrm{Al}}^{+}$ and Ti$_{\mathrm{Al}}^{-}$, (b) Ti$_{\mathrm{Li}}^{+}$ and Ti$_{\mathrm{Li}}^{-}$, and (c) Ti$_{i}^{2+}$ and Ti$_{i}^{2-}$. Large (dark yellow) spheres are Ti, medium (blue) spheres Al, and small (red) spheres H. Only Ti$-$H and Al$-$H bonds with bond lengths smaller than 2 {\AA} are shown.}\label{fig;TiHn}
\end{figure*}

There are drastic changes in the lattice geometry as Ti and Ni impurities are incorporated. The transition-metal impurities tend to pull (AlH$_{4}$)$^{-}$ units closer and/or, in some cases, break Al$-$H bonds to form M$\mathrm{H}_{n}$ complexes. Figure \ref{fig;TiHn} shows the local structures of several substitutional and interstitial Ti configurations in LiAlH$_{4}$. The creation of Ti$_{\mathrm{Al}}^{+}$, for example, leads to the formation of a complex in which the Ti atom is surrounded by $n$=9 hydrogen atoms with the Ti$-$H distances ranging from 1.81 to 1.93 {\AA}, {\it cf.}~Fig.~\ref{fig;TiHn}(a). These Ti$-$H bond lengths are comparable to that in bulk TiH$_{2}$, which is 1.92 {\AA}. Other impurities such as Ti$_{\mathrm{Al}}^{-}$, Ti$_{\mathrm{Li}}^{+}$, Ti$_{\mathrm{Li}}^{-}$, Ti$_{i}^{2+}$, and Ti$_{i}^{2-}$ also lead to formation of TiH$_{n}$ ($n$=6$-$8) complexes with the Ti$-$H distances ranging from 1.78 to 1.95 {\AA}. Similarly, Ni impurities also form NiH$_{n}$ complexes, however the $n$ value is smaller than those of Ti. For example, Ni$_{\mathrm{Al}}^{+}$ is surrounded by 6 hydrogen atoms, instead of 9 hydrogen atoms in the case of Ti$_{\mathrm{Al}}^{+}$. This difference is due to the fact that Ni has fewer empty 3$d$ states than Ti.

Following our argumentation presented previously for other complex hydrides such as NaAlH$_{4}$, LiBH$_{4}$, and Li$_{4}$BN$_{3}$H$_{10}$,~\cite{peles_prb_2007,hoang_prb_2009,wilson-short_prb_2009} the Fermi level of the system is determined by these electrically active impurities if their concentration is larger than that of the native defects. Chances are that this Fermi-level position will be different from the one that would occur in the absence of the impurities. In this case the formation energy of some of the native defects will be reduced. For those impurities that have negative-$U$ character, the Fermi level will simply coincide with the $\epsilon$(+$q$/$-q$) transition level of the impurities.\cite{hoang_prb_2009} For those impurities that exhibit positive $U$ and $\epsilon$(+$q$/0)$<$$\mu_{e}^{\rm int}$$<$$\epsilon$(0/$-q$) such as Ni$_{i}$, {\it cf.}~Fig.~\ref{tab:impurities}, the the impurity have no effect since it prefers the neutral charge state over a range of Fermi-level values that includes the Fermi-level position where the lithium-related defects switch charge state.\cite{hoang_prb_2009} Clearly, the effects of Ni and Ti impurities on shifting the Fermi level depend sensitively on how the impurities are incorporated into LiAlH$_{4}$.

For Ni$_{\rm{Al}}$, Ni$_{\rm{Li}}$, and Ti$_{\rm{Li}}$, the Fermi-level shift will result in lowering the formation energy of $V_{\mathrm{AlH_{4}}}^{+}$. Since the formation and diffusion of $V_{\mathrm{AlH_{4}}}^{+}$ corresponds to formation of H$_{2}$ (and Al) in the mechanism we proposed for the decomposition and dehydrogenation of LiAlH$_{4}$, this may lead to a reduction in the onset decomposition temperature (or the effective decomposition temperature) as seen in experiments. This shift, however, would also result in increasing the formation energy of H$_{i}^{-}$. In this way, Ni$_{\rm{Li}}$, Ni$_{\rm{Al}}$, and Ti$_{\rm{Li}}$ lower the activation energy associated with $V_{\mathrm{AlH_{4}}}^{+}$ but increase that associated with H$_{i}^{-}$. Upon introducing the impurities into LiAlH$_{4}$, $\Delta$ changes from 0 eV up to $-$0.12 eV (Ni$_{\rm{Al}}$), $-$0.44 eV (Ni$_{\rm{Li}}$), or $-$0.42 eV (Ti$_{\rm{Li}}$). Note that, as the Fermi level is shifted, some charged native defects may occur with higher concentrations to counteract the effects of the electrically active impurities, and reduce the magnitude of $\Delta$. When $\Delta$ reaches $-$0.12 eV, the activation energies associated with $V_{\mathrm{AlH_{4}}}^{+}$ and H$_{i}^{-}$ are both equal to 0.99 eV. For lower $\Delta$ values, the activation energy associated with H$_{i}^{-}$ is higher than that associated with $V_{\mathrm{AlH_{4}}}^{+}$. As a result, the process associated with H$_{i}^{-}$ becomes the rate-limiting step in hydrogen desorption. Our analyses, therefore, indicate that Ni$_{\rm{Al}}$, Ni$_{\rm{Li}}$, and Ti$_{\rm{Li}}$ can modify the activation energy. In fact, these impurities can reduce the activation energy by up to about 0.12 eV, which is in good agreement with some experimental observations.~\cite{blanchard_2005,rafiuddin_2010,varin2010928} Using similar arguments, we predict that Ti$_{i}$ would not lower the activation energy for hydrogen desorption according to the proposed mechanism.

\section{\label{sec:sum}Conclusion}

We have carried out a comprehensive first-principles study of native defects and transition-metal impurities in LiAlH$_{4}$. The compound is found to be prone to Frenkel disorder on the Li sublattice. Lithium interstitials and vacancies have low formation energies and are highly mobile, which can participate in lithium-ion conduction, and act as accompanying defects in hydrogen mass transport. We have proposed a specific mechanism for the decomposition of LiAlH$_{4}$ that involves the formation and migration of H$_{i}^{-}$, $V_{\mathrm{AlH_{4}}}^{+}$, and (Li$_{i}^{+}$,$V_{\mathrm{Li}}^{-}$) in the bulk LiAlH$_{4}$. Our calculated activation energy is in agreement with the experimental values. In light of this atomistic mechanism, we are able to explain the decomposition and dehydrogenation of LiAlH$_{4}$, the rate-limiting step in the hydrogen desorption kinetics, and the effects of transition-metal (Ti and Ni) impurities on the onset decomposition temperature and on the activation energy. Our results also suggest that it is the structure of H$_{i}^{-}$ that determines the hydride phase (Li$_{3}$AlH$_{6}$ or LiH) in the decomposition products. This relationship should be further explored in other complex hydrides.

\begin{acknowledgments}
K.~H.~was supported by General Motors Corporation, and A.~J.~by the U.S.~Department of Energy (Grant No. DE-FG02-07ER46434). This work made use of NERSC resources supported by the DOE Office of Science under Contract No.~DE-AC02-05CH11231 and of the CNSI Computing Facility under NSF Grant No.~CHE-0321368. The writing of this paper was partly supported by Naval Research Laboratory through Grant No.~NRL-N00173-08-G001.
\end{acknowledgments}


\end{document}